\documentclass[]{spie}  

\usepackage{amsmath,amsfonts,amssymb}
\usepackage{graphicx}
\usepackage[colorlinks=true, allcolors=blue]{hyperref}

\usepackage[final]{microtype}

\title{Unique MS Lesion Identification from MRI}

\author[1]{Carlos A. Rivas}
\author[1]{Jinwei Zhang}
\author[1]{Shuwen Wei}
\author[2]{Samuel W. Remedios}
\author[1]{\\Aaron Carass}
\author[1,2]{Jerry L. Prince}
\affil[1]{Image Analysis and Communications Laboratory, Dept.~of~Electrical~and~Computer~Engineering, Johns~Hopkins~University,~Baltimore,~MD~21218,~USA\vspace*{0.4em}}
\affil[2]{Dept. of Computer Science, Johns Hopkins University, Baltimore, MD 21218, USA}

\authorinfo{Further author information: (Send correspondence to C.R.)\\C.R.: E-mail: crivas6@jh.edu, Telephone: +1 253 548 6615}

\begin{document} 
\maketitle

\begin{abstract}
Unique identification of multiple sclerosis~(MS) white matter lesions~(WMLs) is important to help characterize MS progression.
WMLs are routinely identified from magnetic resonance images~(MRIs) but the resultant total lesion load does not correlate well with EDSS\footnote[1]{Expanded disability status scale~(EDSS) correlates with disability.}; whereas mean unique lesion volume has been shown to correlate with EDSS.
Our approach builds on prior work by incorporating Hessian matrix computation from lesion probability maps before using the random walker algorithm to estimate the volume of each unique lesion.
Synthetic images demonstrate our ability to accurately count the number of lesions present.
%
%
The takeaways, are: 1)~that our method correctly identifies all lesions including many that are missed by previous methods; 2)~we can better separate confluent lesions; and 3)~we can accurately capture the total volume of WMLs in a given probability map.
This work will allow new more meaningful statistics to be computed from WMLs in brain MRIs.
\end{abstract}



\section{INTRODUCTION}
\label{s:intro}  
Multiple sclerosis~(MS) is a neurodegenerative disease that is characterized by white matter lesions~(WMLs) that are readily identifiable~\cite{zhang2019miccai, zhang2024spie} in T2-weighted~(T2w) fluid-attenuated inversion recovery~(FLAIR) magnetic resonance imaging~(MRI) of the head.
The total volume, also known as \textit{lesion load}, of these WMLs was for a long time considered a useful biomarker of disease; however lesion load has never been shown to correlate well with the expanded disability status scale~(EDSS)~\cite{vanwalderveen1995n, schreiber2001ans}.
This is troublesome as the EDSS is considered a useful clinical marker of disability and disease progression in MS.
Some recent work has suggested that the mean volume per lesion may correlate with EDSS~\cite{dworkin2018automated}, moreover there is increasing interest in identifying both new lesions and tracking the progression of a lesion through the time course of disease to offer more insight into MS.
Simple connected component analysis typically identifies hundreds of lesions, which is impossible to confirm manually and such connected component analyses cannot distinguish two unique lesions that are touching.
Thus, there is a need for a fully automated unique lesion identification algorithm.

The unique identification of WMLs is difficult and since there are currently no available public databases, it is not feasible to train a deep network approach for this problem.
Thus, we build upon the work of Dworkin~\textit{et~al.}~\cite{dworkin2018automated} to develop a unique WML identification algorithm that also estimates the volume of each unique lesion.
Wynen~\textit{et~al.}~\cite{wynen2024isbi} have proposed a deep network approach for handling confluent lesions, however their method is not currently available for comparison.
We present synthetic data to demonstrate that our approach is capturing lesions that are missed by Dworkin \textit{et al.} and experiments on real data that highlight the benefits of our approach.

\section{METHOD}
\label{s:section}
\begin{figure}[!tb]
\begin{center}
\begin{tabular}{@{}c@{}c c @{}c@{}c c @{}c@{}c}

\includegraphics[trim=95 50 243 100, clip, height=3.0cm]{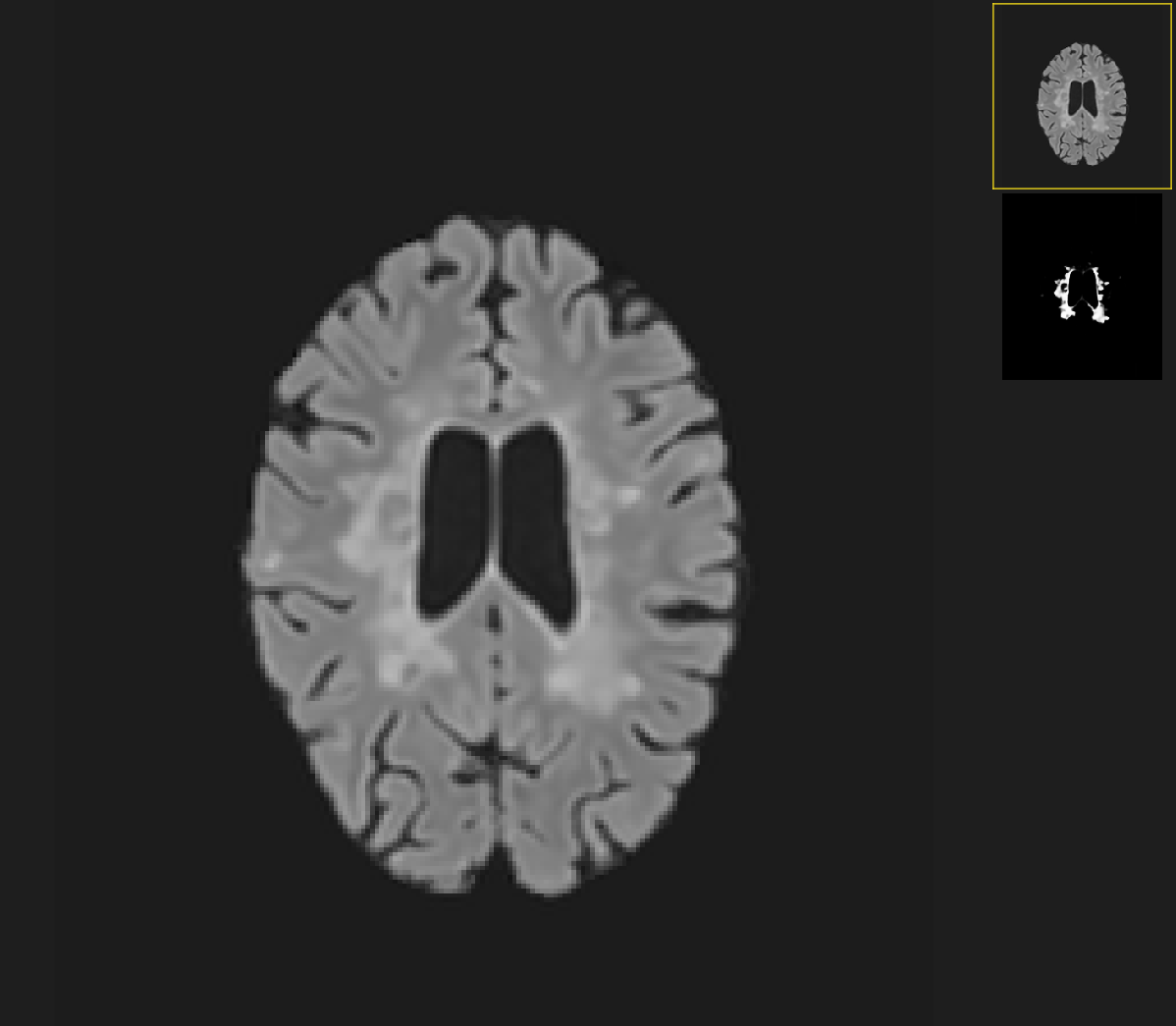} &
\includegraphics[trim=95 50 243 100, clip, height=3.0cm]{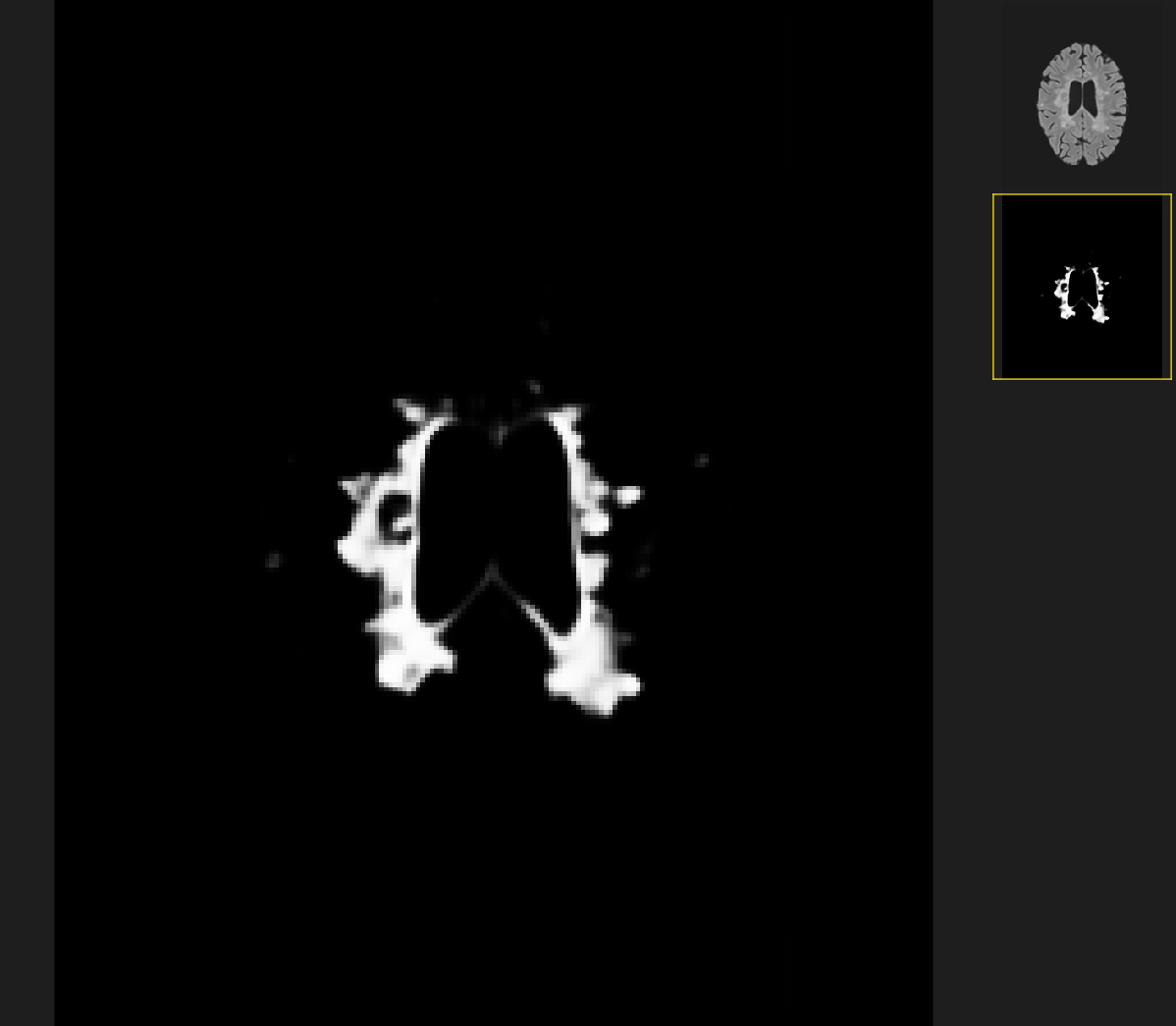} &&
\includegraphics[trim=95 50 243 100, clip, height=3.0cm]{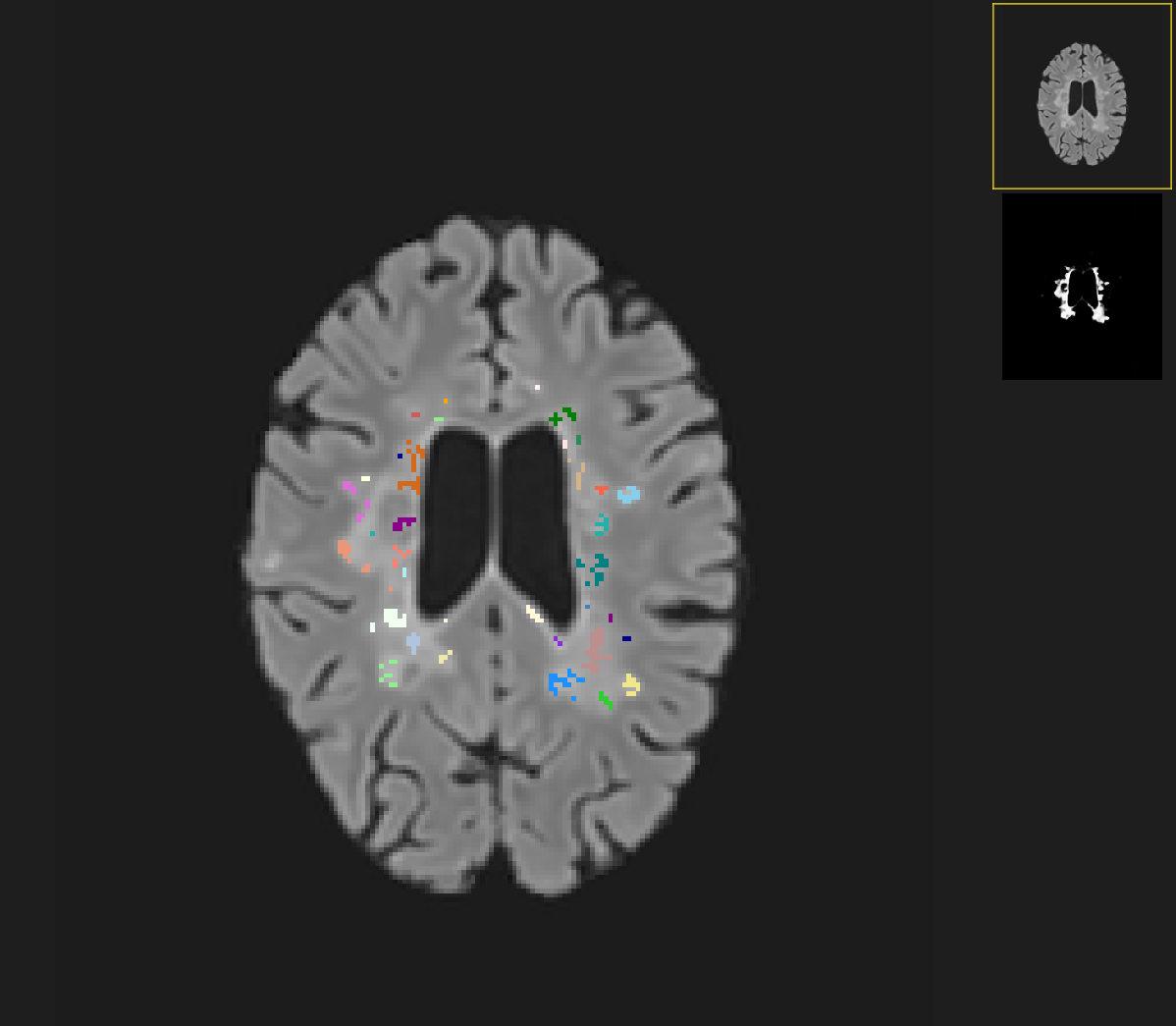} &
\includegraphics[trim=95 50 243 100, clip, height=3.0cm]{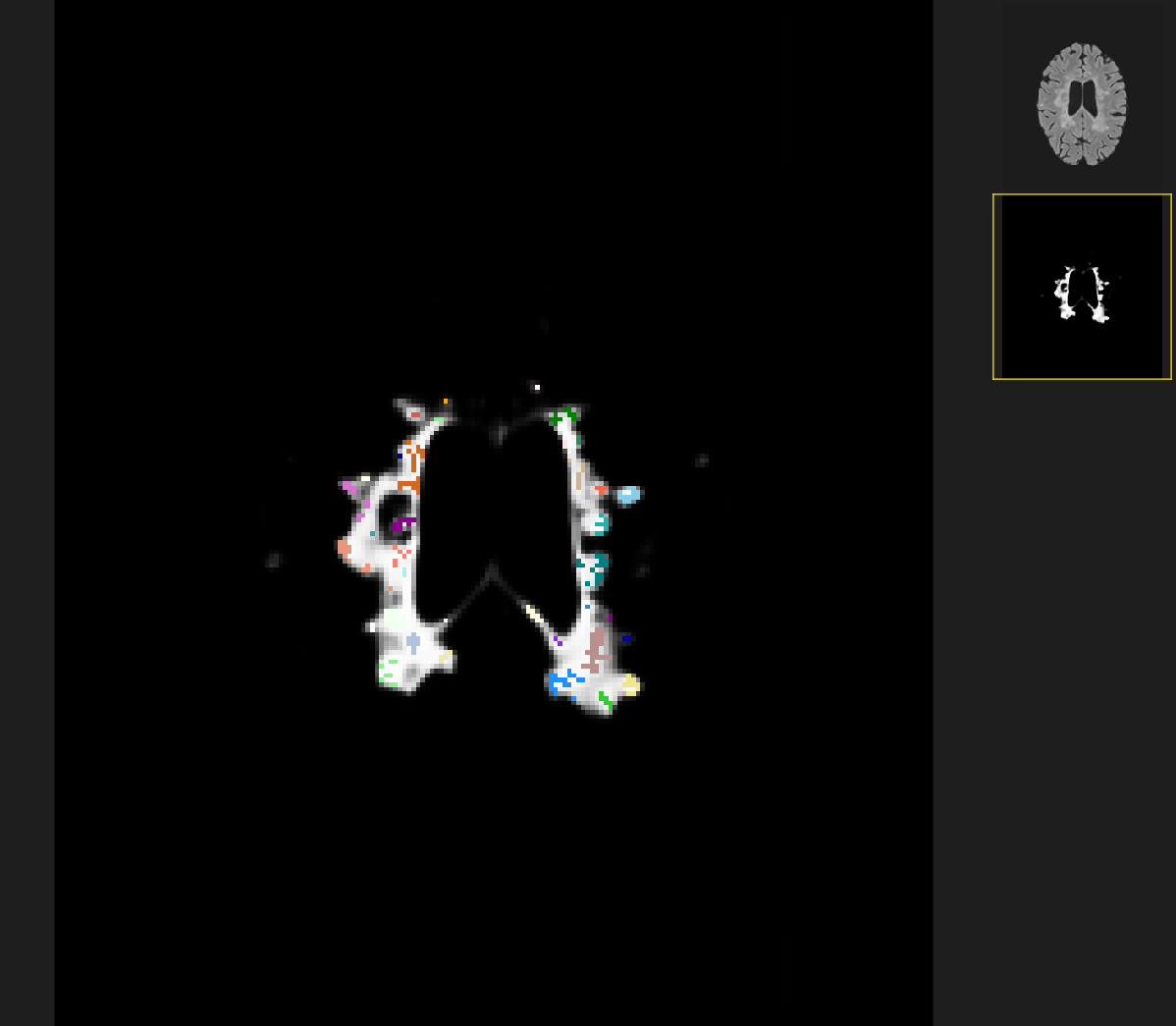} &&
\includegraphics[trim=95 50 243 100, clip, height=3.0cm]{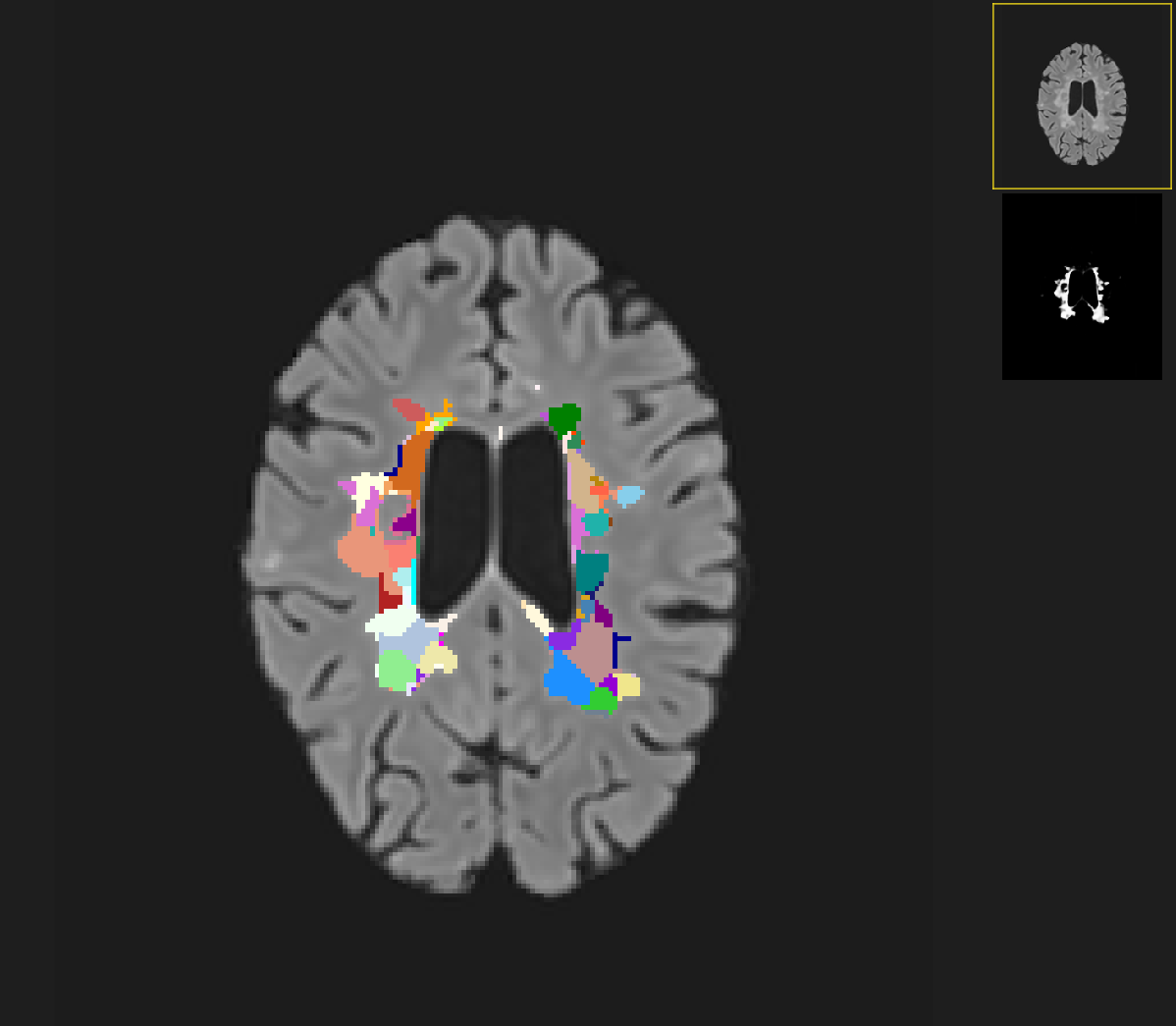} &
\includegraphics[trim=95 50 243 100, clip, height=3.0cm]{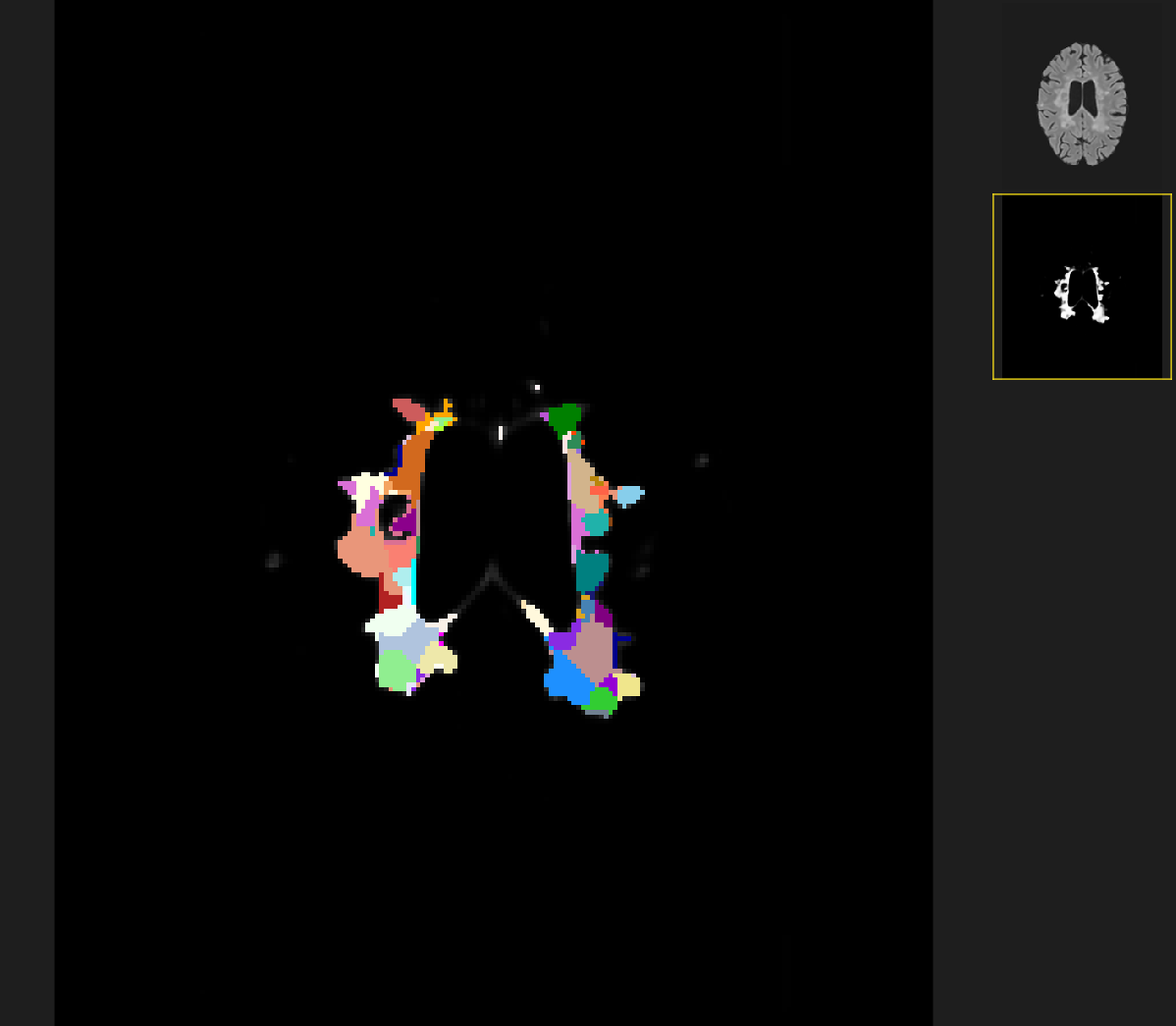} \\
\multicolumn{2}{c}{\textbf{(a)}} && \multicolumn{2}{c}{\textbf{(b)}} && \multicolumn{2}{c}{\textbf{(c)}}\\

\end{tabular}
\end{center}
\caption{
\textbf{(a)}~Axial FLAIR image and its corresponding lesion probability map~\cite{zhang2024isbi} for an MS subject. 
\textbf{(b)}~The same images as (a) with the centers of the unique lesions indicated by a color overlay. 
\textbf{(c)}~The unique lesions after our growth step.
}
\label{f:pipeline}
\end{figure}

Figure~\ref{f:pipeline} provides an overview of the processing steps that we describe below.
In brief, these are preprocessing, lesion center identification, and lesion growth to assign lesioned tissue to a unique lesion.

\paragraph{Dataset and Preprocessing}
We use data from the publicly available 2016 MICCAI MS segmentation challenge dataset~\cite{commowick2018objective}.
This dataset contains 15 MS subjects, with images acquired from three different scanners, and includes five contrast-weighted images: T1w, T1w with gadolinium, T2w, PDw, and FLAIR.
The multi-contrast images of each subject were denoised using a non-local means algorithm, followed by rigid body registration to the FLAIR contrast, skull stripping, and N4 bias field correction.
A lesion probability map, was generated using SELF~\cite{zhang2024isbi}.
An example FLAIR image with the corresponding lesion probability map is shown in Fig.~\ref{f:pipeline}(a).

\paragraph{Lesion Center Identification}
We build upon the work of Dworkin~\textit{et~al.}~\cite{dworkin2018automated}.
We calculate the Hessian matrix for each voxel of the 
SELF lesion probability map and their corresponding eigenvalues.
We use the eigenvalues to identify local maxima of the probability map as negative definite Hessian matrices; these local maxima are presumed to correspond to the centers of lesions.
We exclude local maxima below a lesion probability threshold, as these are presumed to be noisy lesions centers arising from low probability lesion areas.
We use connected components analysis with 26-connectivity to identify groups of local maxima, which represent plateaus at a lesion center.
This results in an image with identified lesions centers and allows us to produce a total lesion count. An example of identified lesion seeds is shown as a color overlay in Fig.~\ref{f:pipeline}(b).

\paragraph{Lesion Center Growth}
With the lesion centers identified, we use the Random Walker~\cite{grady2006pami} algorithm to associate lesioned tissue uniquely with lesion centers.
To do this, we use our identified lesion centers as seeds and the SELF lesion probability map as underlying image.
We threshold the SELF lesion probability map to remove low-probability lesion values, reducing the potential for erroneous assignments. 
Values below the threshold are set to a negative value to ensure that the Random Walker algorithm excludes them from being associated with a lesion center.
An example of the uniquely identified lesions is shown as a color overlay in Fig.~\ref{f:pipeline}(c).



\paragraph{Program Parameters}
Complete details on parameter fine tuning will be included in the final manuscript.

\section{EXPERIMENTS and RESULTS}


%
Lesion count accuracy was assessed using synthetically generated images, evaluating the ability to detect small, low-probability, and confluent lesions.
Additionally, counts were performed on the 15 subjects described above and compared both quantitatively and qualitatively to the results from a binary threshold probability map (as a baseline comparison), Dworkin's method~\cite{dworkin2018automated}, and our proposed approach.
Lesion growth was assessed by comparing the total volume of summed lesions to the volume obtained from the thresholded binary map and to the total summed lesion volume from Dworkin's method.

\begin{figure}[!tb]
\begin{center}
\begin{tabular}{@{}c@{\hspace{0.5cm}}c@{\hspace{0.5cm}}c@{\hspace{0.5cm}}c@{}} 
\textbf{\hspace*{3ex}Synthetic Image} & \textbf{Binary Threshold} & \textbf{Dworkin Method} & \textbf{Our Method} \\
\raisebox{-0.45cm}{\includegraphics[trim=9 9 9 0, clip, height=4cm]{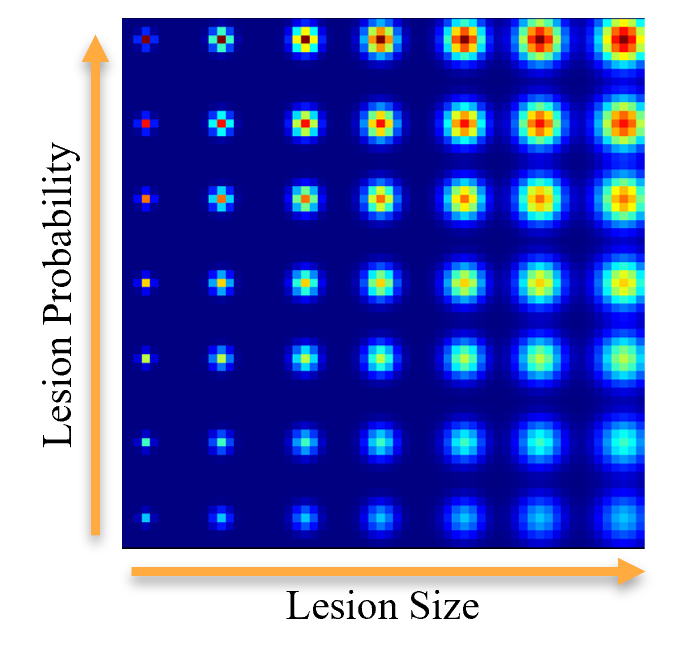}} &
\reflectbox{\includegraphics[trim=12 0 12 0, clip, width=3.5cm]{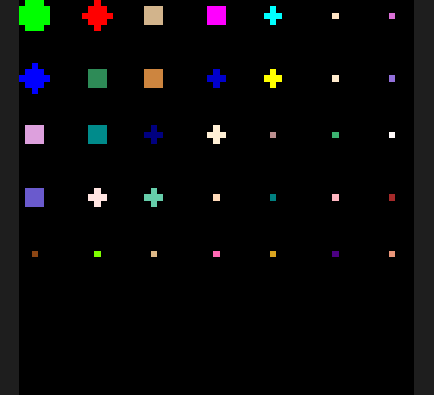}} &
\reflectbox{\includegraphics[trim=12 0 12 0, clip, width=3.5cm]{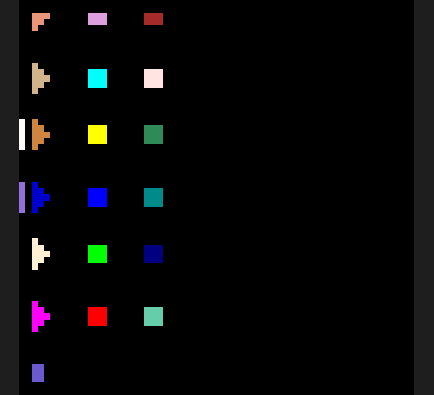}} &
\reflectbox{\includegraphics[trim=12 0 12 0, clip, width=3.5cm]{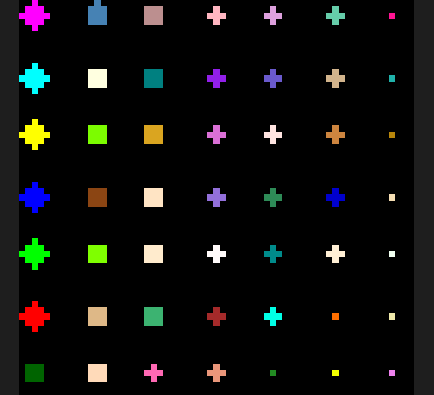}} \\

\end{tabular}
\end{center}
\caption{From left to right: A slice of a synthetic lesion probability map without lesion confluence displayed with the jet colormap~(Synthetic Image), the connected components from a binary threshold approach~(Binary Threshold), the output of the Dworkin Method, and finally, our approach.}
\label{f:synthetic}
\end{figure}

\begin{figure}[!tb]
\begin{center}
\begin{tabular}{@{}c@{\hspace{0.05cm}}c@{\hspace{0.05cm}}c@{\hspace{0.05cm}}c@{\hspace{0.05cm}}c@{}} 
\textbf{Synthetic Image} & \textbf{Manual} & \textbf{Binary} & \textbf{Dworkin Method} & \textbf{Our Method} \\

\includegraphics[trim=150 175 260 152, clip, width=3.3cm]{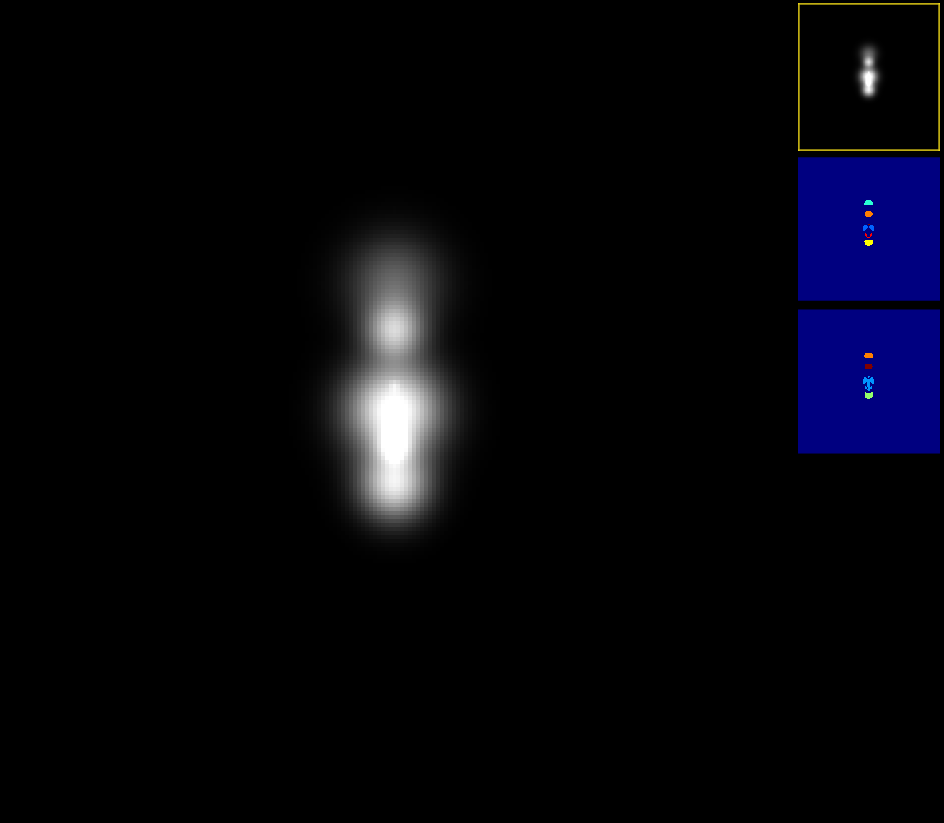} &
\includegraphics[trim=15 20 15 13.25, clip, width=3.3cm]{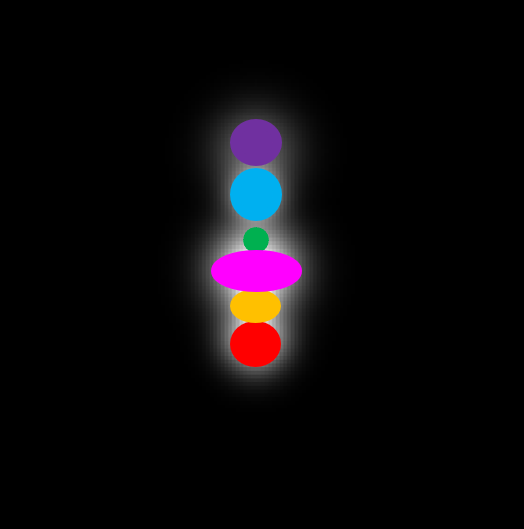} &
\includegraphics[trim=75 70 75 55, clip, width=3.3cm]{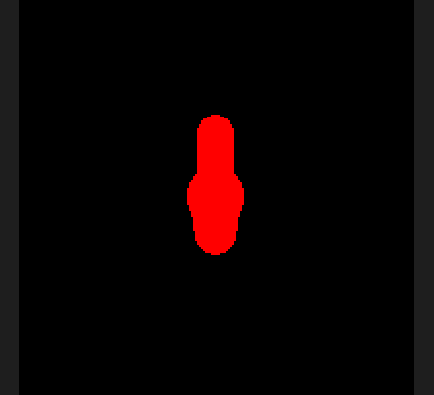} &
\includegraphics[trim=75 70 75 55, clip, width=3.3cm]{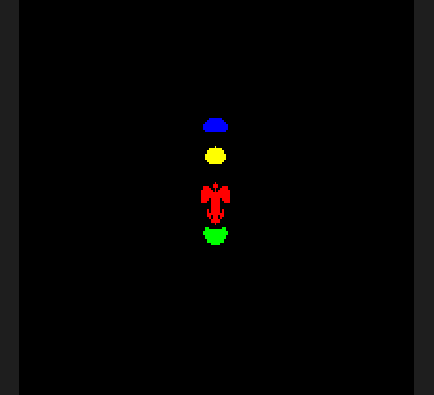} &
\includegraphics[trim=75 70 75 55, clip, width=3.3cm]{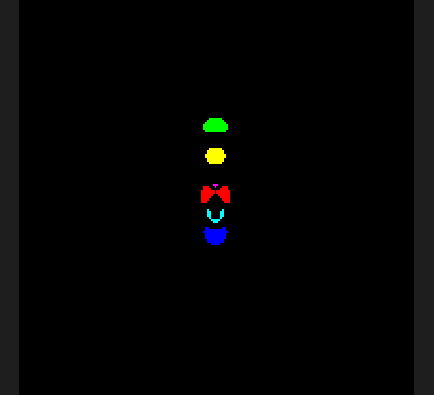} \\

\end{tabular}
\end{center}
\caption{From left to right: A slice of a synthetic image with six confluent lesions which are manually identified~(Manual) and the results of identification via binary threshold~(Binary), the Dworking Method, and our proposed method. Binary thresholding identified one lesion, Dworkin identified four lesions, whereas our method identified all six lesion centers.}
\label{f:confluent}
\end{figure}

\paragraph{Synthetic Lesion Data}
Fig.~\ref{f:synthetic} shows our synthetic lesion probability map, as well as the results of using a binary threshold to identify lesions, the Dworkin Method, and our proposed method.
The synthetic image shows the two anticipated modes of failure in uniquely identifying lesions: 1)~lesion probability and 2)~lesion size.
Lesion probability affects both the binary threshold approach and the Dworkin method.
While lesion size affects only the Dworkin method, with small lesions not being identified.
A surprising failure of the Dworkin method is the double counting of lesions.
Fig.~\ref{f:confluent} shows another synthetic, but more realistic, lesion probability map with six confluent lesions present in the image.
The Dworkin method merged two lesions and missed another lesion, resulting in a lesion count of four, whereas our approach correctly identified six unique lesions.
These synthetic experiments reveal the drawbacks of the Dworkin method, including inaccuracies related to missing small lesions, issues with low-probability lesions, and occasional double counting of individual lesions, which are not encountered in our proposed method.

\begin{figure}[!tb]
\begin{center}
\begin{tabular}{c c c}
\includegraphics[width = 0.55\textwidth]{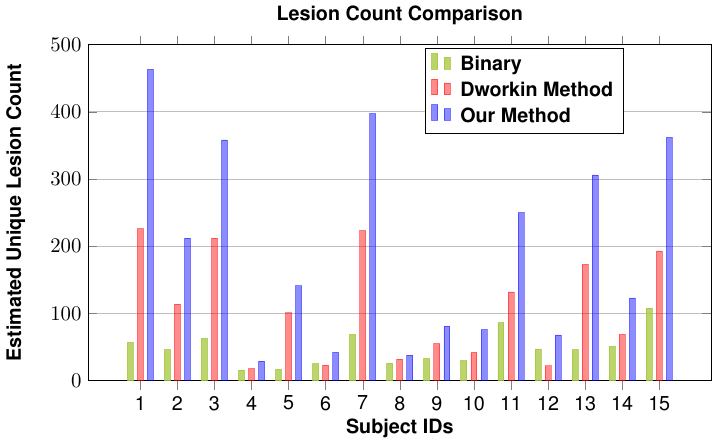} &&
\includegraphics[width = 0.38\textwidth]{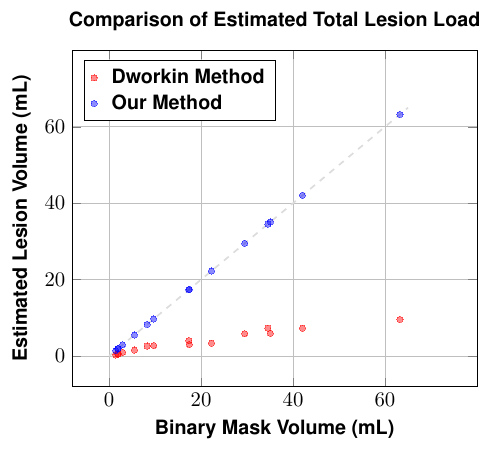}\\
\textbf{(a)} && \textbf{(b)}\\[-.5em]
\end{tabular}
\end{center}
\caption{\textbf{(a)}~Lesion counts from Binary threshold with connected components, the Dworkin method, and our method.
\textbf{(b)}~Total volume of captured lesions from binary thresholding, the Dworkin method, and after the growth phase of our method.}
\label{f:plots} 
\end{figure}

Figure~\ref{f:plots}(a) shows a comparison of the lesion counts for the 15 subjects, using the connected component analysis from binary thresholding approach, the Dworkin method, and our proposed method.
For every subject tested, our counting method identified more lesions than the other two methods.
Specifically, for subjects 6 and 12, the Dworkin method underperformed compared to the connected component analysis from binary thresholding.
This further highlights the innacuracies of the Dworkin method.
Fig.~\ref{f:plots}(b) shows a comparison of the total volume of captured lesions.
Our method matched the volume of the binary mask run at the same threshold level due to the design of our lesion growth implementation, which fills the area of the lesion probability map it receives as input.
In contrast, the Dworkin method clearly underestimated the total lesion volume.

\begin{figure}[!tb]
\begin{center}
\begin{tabular}{c c c} 
\includegraphics[trim=330 8 0 5, clip, height=5cm]{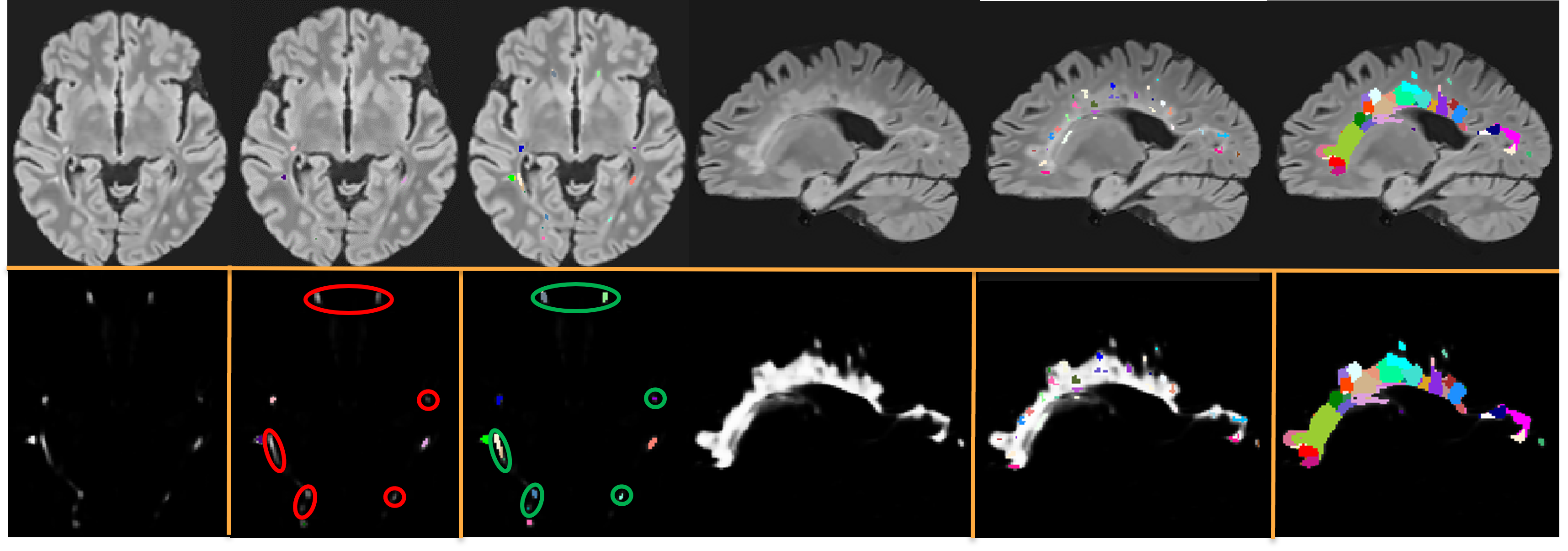} && \includegraphics[trim=5 8 418 0, clip, height=5cm]{Sub7_Sub9.png} \\
\textbf{(a)} && \textbf{(b)}\\
\end{tabular}
\end{center}
\caption{
\textbf{(a)}~On the left is the sagittal MR image displayed with the lesion probability map above it. 
The middle images show the Dworkin-identified lesions overlaid, and the right images display the results of our method. 
\textbf{(b)} Same display order as (a), but on the axial MR image of another subject. 
The red-circled areas highlight lesions that the Dworkin method missed, while the green-circled areas show lesions that our method identified.}
\label{f:results}
\end{figure}

\paragraph{Real Lesion Data}
Figures~\ref{f:results}(a) and~(b) show the outputs of the Dworkin method and our proposed method side by side on sagittal and axial images for two subjects, along with the lesion probability maps.
Fig.~\ref{f:results}(a) shows a subject with the second highest number of lesions among all 15 subjects in our experiments.
In Fig.~\ref{f:results}(a), the stark differences between the complete processing of our method versus the Dworkin method is revealed, highlighting the importance of the growth step.
While the Dworkin method did an admirable job detecting lesions, the total volume it captured was only a small subset of what was captured by the lesion probability map.
With our additional growth step, individual lesions in the image were filled to the extent of the binary threshold.
Fig.~\ref{f:results}(b) shows a subject with much fewer lesions and smaller lesion sizes compared to Fig.~\ref{f:results}(a).
In Fig.~\ref{f:results}(b), the Dworkin method clearly missed many lesions~(indicated by the red circles), whereas our method accurately detected those same lesions~(indicated by the green circles).

\section{DISCUSSION AND CONCLUSION}
In this work, we demonstrate the effectiveness of our unique lesion identification method compared to the previously reported method.
Our method successfully identifies small and low-probability lesions, which are two areas where the previous method fails. 
Additionally, our method can detect unique lesions in complex cases involving confluent lesions.
Future work includes applying the proposed method to a larger cohort of MS patients and exploring the longitudinal tracking of individual lesions.
Future work also includes verifying the accuracy of lesion counts with a trained radiologist.

\centerline{\textbf{This work has not been submitted for publication or presentation elsewhere.}}

\section*{Acknowledgments}
This work was partially supported by the NIH under NEI grant R01-EY032284 (PI: J.L. Prince) and by the National Science Foundation Graduate Research Fellowship under Grant No. DGE-1746891 (Remedios).



\bibliography{references.bib} 
\bibliographystyle{spiebib} 

\end{document}